\documentclass[aps,pre,10pt,twocolumn,superscriptaddress]{revtex4-2}
\usepackage{graphicx}
\usepackage{subcaption}
\usepackage{epstopdf}
\usepackage{xcolor}
\usepackage{amsmath}
\usepackage{amssymb}
\usepackage[utf8]{inputenc}
\usepackage{graphicx,float,hyperref}
\usepackage{natbib}
\usepackage{graphicx,latexsym}

\begin{document}

\title{A Unified Protocol for Arbitrary $Z(\theta)$ Rotations of a Class of Bosonic cQED Qubits}
\author{Vahid Shaghaghi}
\email{Vahid.physics@gmail.com}
\affiliation{Department of Optics, Palacky University, 17. listopadu 1192/12, 77146 Olomouc, Czechia}

\begin{abstract}
Bosonic-qubit encodings provide diverse approaches to storing and processing quantum information, with logical operations often implemented through encoding-specific physical strategies. Here, we identify a class of bosonic Fock-based qubits in circuit QED for which distinct implementations of arbitrary $Z(\theta)$ rotations share a common structure. The applicability of the proposed framework is determined by specific conditions involving the greatest common divisor of the photon numbers and a parity criterion for the logical codewords. These conditions provide a systematic means of identifying bosonic Fock-based cQED qubits that admit the common implementation.
\end{abstract}


\maketitle 


\section{INTRODUCTION}

Breakthrough theoretical results \cite{Deutsch1992,Grover1996,Bennett1997,Shor1999}, which show that certain classical computations can be significantly shortened in time by quantum computations, have inspired a variety of proposals—trapped ions \cite{Cirac1995,Leibfried2003,Haffner2008,Blatt2012} and neutral atoms \cite{Bloch2008,Saffman2010, Henriet2020,Kaufman2021} , just to name a few—with the aim of achieving such supremacy.

Superconductors, as well-established efficient hardware, have impressively proven their effectiveness in practice. The implementation of quantum optical ideas in superconductor-based electrical circuits has led to the flourishing area of circuit QED \cite{Blais2021}. Subsequently, the creation of artificial atoms with long coherence times and good controllability have formed the innovative idea of quantum information processing in circuit QED\cite{Blais2004}. In this realm, superconducting two-level \cite{Vion2002,Koch2007,Schreier2008} and multi-level systems \cite{Leghtas2013,Terhal2016,Ofek2016} have been introduced as the fundamental units of quantum information, known as qubits.  

 Among these qubits, bosonic cQED qubits have shown remarkable features such as single photon loss as the dominant error \cite{Ying2015}, long coherence time \cite{Reagor2016}, negligible intrinsic dephasing \cite{Rosenblum2018}, surpassing the break-even point \cite{Sivak2023}, and dealing with only a single physical system (a mode) - have made them potential candidates for practical quantum computations. Consequently, developing flexible and systematic methods for controlling these qubits is therefore essential for advancing bosonic quantum information processing and realizing scalable quantum technologies.
 
 Logical operations on these qubits are often formulated in an encoding-specific manner, with apparently different physical procedures used to realize the same logical operation \cite{Mirrahimi2014,Heeres2017,Hu2019,Ma2020,Xu2022,Teoh2023,Marquet2024,Shaghaghi,Gautier2023, Xu2023}, just to name a few. This diversity of implementations raises a natural question: whether seemingly distinct realizations of a logical operation can be understood within a common framework.
 
 Here, we address this question for arbitrary $Z(\theta)$ rotations in bosonic Fock-based qubits. Arbitrary $Z(\theta)$ rotations are essential to quantum computing because they enable precise control of the relative phase between computational basis states, providing a fundamental resource  for manipulating quantum information. Together with rotations about other axes, they enable the implementation of universal single-qubit gates and contribute to the construction of arbitrary quantum circuits. Consequently, developing systematic and broadly applicable protocols for arbitrary $Z(\theta)$ rotations can facilitate quantum gate implementation across different bosonic Fock-based encodings and expand the available toolbox for cQED quantum processors\cite{Liu2026}.
 
  To this end, we employ the protocol introduced for implementing a logical $ZZ(\theta)$ gate in a dual-rail qubit \cite{Teoh2023}. Although we use the same protocol, the key difference lies in how the relative $\pi$ phase between the codewords is imparted: they use a joint parity unitary to implement a $\pi$ phase shift for a specific encoding, whereas our approach employs a parity criterion arising from the rescaling of photon numbers by the greatest common divisor, which determines the dispersive interaction time. This distinction enables us to identify a class of Fock-based cQED qubit admitting a common implementation strategy. More broadly, we provide a systematic way to recognize when a bosonic Fock-based cQED qubit can benifit from the same implementation framework.  
 
 To illustrate the scope of this framework, we consider five bosonic Fock-based qubits examples in which $Z(\theta)$ rotations can be implemented using different physical approaches. The examples illustrate both the success and failure of the proposed protocol in the implementing $Z(\theta)$ rotations within a common framework, thereby demonstrating evidence for the generality of the proposed approach.
 
  The paper is organized as follows. First, we present the two essential interactions underlying the protocol: the dispersive and Rabi interactions, and show how the greatest common divisor of the photon numbers can be used to generate the required  $\pi$-phase shift. Then, we demonstrate how these interactions are combined to construct a six-step protocol for the implementation of arbitrary $Z(\theta)$ rotations. Next, we establish the conditions that identify the class of bosonic cQED qubit amenable to this common implementation framework. Finally, we demonstrate the scope and applicability of the protocol through several examples of bosonic Fock-based qubits.

\section{Architecture}

We consider a cQED architecture comprising a two-level ancilla  
\begin{equation}
|\phi\rangle=\alpha|g\rangle+\beta|e\rangle
\end{equation}
dispersively coupled to a bosonic mode
\begin{equation}
|\psi\rangle=\eta|0\rangle_L+\gamma|1\rangle_L
\end{equation}
where
\begin{equation}
 \qquad |0\rangle_L=\sum_{n\in S_0}^{}c_n|n\rangle,\qquad |1\rangle_L=\sum_{m\in S_1}^{}c_m|m\rangle
\end{equation}
$\alpha$, $\beta$, $\eta$, $\gamma$, $c_n$, and $c_m$ could be complex numbers. $|g\rangle$ and $|e\rangle$ represent ground and excited states of the ancilla, respectively. And
\begin{align}
S_0=\{n:c_n\neq 0\}, \qquad S_1=\{m:c_m\neq 0\}
\end{align}
 the governing interaction Hamiltonian is
\begin{equation}\label{key}
\hat{H}_{dis}=\chi |e\rangle\langle e| \hat{a}^\dagger\hat{a}
\end{equation}
$\chi$ is the dispersive strength, $ \hat{a}^\dagger$ and $\hat{a}$ are creation and annihilation operators for the bosonic cQED qubit. Consequently, the dispersive unitary operator is represented as

\begin{equation}\label{key}
\hat{U}_{dis}(t)=e^{-it\hat{H}_{dis}}=
\begin{bmatrix}
e^{-i\chi \hat{a}^\dagger\hat{a}t} & 0 \\
0 & 1
\end{bmatrix}
\end{equation}
it follows from the unitary that when $t_{\pi}=\frac{\pi}{k\chi}$, and the ancilla is in the excited state, the resulting unitary operation corresponds to a $\pi$-phase shift operation. The parameter $k$ plays a key role in distinguishing our approach from \cite{Teoh2023}. The reason behind the appearance of the parameter $k$ in the denominator is to rescale the photon numbers of the codewords into even and odd integers, thereby enabling the application of the parity criterion to implement a relative $\pi$ phase between the codewords.  

$k$ is obtained from the greatest common divisor (gcd) of the union of sets $S_0$ and $S_1$
\begin{equation}
k=\gcd(S_0\cup S_1)
\end{equation}
thereby, $n$ and $m$ are multiples of $k$
\begin{align}
n=kn',\qquad m=km'
\end{align}
To ensure that every occupied Fock component within each logical state acquires the same phase, and the two logical states acquire opposite phases, we employ the parity criterion as
\begin{multline}
\\n' \equiv p\quad (\textrm{mod} 2)\quad \forall n\in S_0\\
m' \equiv p+1\quad (\textrm{mod} 2)\quad \forall m\in S_1\\
\end{multline}
where $p\in \{0,1\}$. If  each component within the same logical codeword obtain a different phase, the dispersive interaction can modify the encoded state rather than simply apply a logical phase.

The dispersive interaction applied for time $t_{\pi}=\frac{\pi}{k\chi}$ implements a logical conditional phase operation equivalent to a $\pi$ phase difference between the codewords. Straightforwardly, we can reach this result as follows
\begin{equation}
U_{dis}(t_{\pi})|\psi\rangle\propto \eta\sum_{n\in S_0}^{}c_ne^{-ip\pi}|n\rangle+\gamma\sum_{m\in S_1}^{}c_me^{-i(p+1)\pi}|m\rangle
\end{equation}
with $p=0$, that is equivalent to
\begin{equation}
U_{dis}(t_{\pi})|\psi\rangle\propto |0\rangle_L-|1\rangle_L
\end{equation}
 one logical basis state remains intact, while the other logical basis state gets a $\pi$-phase shift.\\
 
The main point is that the gcd identifies a useful common scale; the parity condition establishes whether the resulting interaction implements the required logical operation.
 
Another necessary interaction is Rabi interaction, governing the rotations of the ancilla
\begin{equation}\label{key}
\hat{H}_{Rabi}=\Omega_d(t)[e^{-i\phi}\sigma_++e^{i\phi}\sigma_-]
\end{equation}
where $\Omega_d(t)$ is the drive amplitude and $\phi$ defines the rotation axis. $\sigma_+$ and $\sigma_-$ are raising and lowering operators for the ancilla. Therefore, the time-evolution operator becomes
\begin{equation}\label{key}
\hat{U}_{Rabi}(t)=e^{-it\hat{H}_{Rabi}}=
\begin{bmatrix}
\cos(\Omega_d t) & -i e^{-i\phi}\sin(\Omega_d t) \\
-i e^{i\phi}\sin(\Omega_d t) & \cos(\Omega_d t)
\end{bmatrix}
\end{equation}
By appropriate controlling and sequencing these two interactions, the architecture provides the physical basis for the following protocol to implement arbitrary $Z(\theta)$ rotations on bosonic Fock-based qubits.


\section{Protocol}

The protocol consists of six physical operations described as follows: by appropriately choosing the interaction parameters and durations at each step, the protocol engineers the desired relative phase between the logical states while preserving the encoded qubit.

$U_1$: the ancilla is initialized in the ground state
\begin{equation}
|\Psi\rangle\propto(|0\rangle_L+|1\rangle_L)|g\rangle
\end{equation}
$U_2$: a $\frac{\pi}{4}$-rotation places the ancilla in a superposition state
\begin{equation}
|\Psi\rangle\propto|0\rangle_L(|g\rangle+|e\rangle)+|1\rangle_L(|g\rangle+|e\rangle)
\end{equation}
$U_3$: the dispersive interaction imparts a $\pi$-phase to the state $|1\rangle_L$ when the ancilla is in the excited state,
\begin{equation}
|\Psi\rangle\propto|0\rangle_L(|g\rangle+|e\rangle)+|1\rangle_L(|g\rangle-|e\rangle)
\end{equation}
$U_4$: $\frac{\theta}{2}$ ancilla rotation gives rise to
\begin{equation}
|\Psi\rangle\propto e^{-i\theta/2}|0\rangle_L(|g\rangle+|e\rangle)+e^{i\theta/2}|1\rangle_L(|g\rangle-|e\rangle)
\end{equation}
$U_5$: the second $\pi$-phase shift disentangles the bosonic cQED qubit from the ancilla
\begin{equation}
|\Psi\rangle\propto (e^{-i\theta/2}|0\rangle_L+e^{i\theta/2}|1\rangle_L)(|g\rangle+|e\rangle)
\end{equation}
$U_6$: a $\frac{\pi}{4}$ ancilla rotation returns the ancilla to its ground state
\begin{equation}
|\Psi\rangle\propto e^{-i\theta/2}(|0\rangle_L+e^{i\theta}|1\rangle_L)|g\rangle
\end{equation}
this result demonstrates that the bosonic cQED qubit undergoes a $Z(\theta)$ rotation up to a global phase factor $e^{-i\theta/2}$
\begin{equation}
U_6U_5U_4U_3U_2U_1|\Psi\rangle\propto e^{-i\theta/2}Z(\theta)|\Psi\rangle
\end{equation}

Main point is that this protocol is possible only for bosonic cQED qubits that satisfy the odd condition for $m'-n'$, which is equivalent to the parity criterion. Therefore,  We can classify bosonic cQED qubits that their $Z(\theta)$ rotation can be implemented by this protocol
\begin{multline}
\mathcal{C}_z=\{\text{span}\{|0\rangle_L,|1\rangle_L\}, k=\gcd(S_0\cup S_1),\\ n/k=p\quad (\text{mod} 2)\forall n\in S_0,\\
 m/k=p+1, \quad (\text{mod} 2)\forall m\in S_1 \}\\
\end{multline}  
the significance of this class is that all bosonic Fock-based cQED qubit encodings belonging to it admit a common procedure for implementing arbitrary rotations about the $Z$ axis. Consequently, rather than designing a separate gate sequence for each encoding, we can construct a single protocol whose implementation depends only on the defining properties of the class.


\section{EXAMPLES}

To demonstrate the applicability and generality of the protocol, we consider several representative examples of bosonic Fock-based qubits, thereby illustrating that the protocol is not restricted to a particular encoding scheme.
\subsection{Cat Qubit}
Assuming $|\alpha\rangle$ is a coherent state for a single bosonic mode, two orthogonal states (cat states) define the basis states for a single logical qubit \cite{Cochrane1999}
\begin{equation}
|0\rangle_L=\mathcal{N_+}(|\alpha\rangle+|-\alpha\rangle),\quad |1\rangle_L=\mathcal{N_-}(|\alpha\rangle-|-\alpha\rangle)
\end{equation}
where the normalization constants are given by
\begin{equation}
\mathcal{N}_{\pm}=(2\pm2e^{-2|\alpha|^2})^{-1/2}
\end{equation}
$|0\rangle_L$ only contains even Fock state terms, while $|1\rangle_L$ contains odd Fock state terms.
The $Z(\theta)$ rotation of this qubit has been implemented by chirping the pump frequency\cite{Iyama2024}.

This rotation can also be implemented by the protocol. Furthermore, the greatest common divisor is
\begin{equation}
k=\gcd\{\text{even, odd}\}=1
\end{equation}
Since $|0\rangle_L (|1\rangle_L)$ contains even (odd) parity, the qubit belongs to $\mathcal{C}_z$.
Therefore, the required dispersive interaction time to impart a desired phase on the logical qubit is given by
\begin{equation}
t_{\pi}=f(\chi,k)=\frac{\pi}{\chi}
\end{equation}

\subsection{Binomial Qubit}
A well-known and simple binomial qubit is defined by the following codewords \cite{Michael2016}
\begin{equation}
|0\rangle_L=\frac{1}{\sqrt{2}}(|0\rangle+|4\rangle),\quad |1\rangle_L=|2\rangle
\end{equation}
This qubit has been explored in cQED platform, which optimized GRAPE control pulses have been used to implement the $Z(\theta)$ rotation\cite{Hu2019}.

For this qubit, the greatest common divisor becomes
\begin{equation}
k=\gcd\{0,2,4\}=2
\end{equation}
we can rewrite the photon numbers as the multiples of $k$, $0=0\times k$, $2=1\times k$, and $4=2\times k$. Rescaling the codewords leads to even parity for $|0\rangle_L$ and odd parity for $|1\rangle_L$; So, the binomial code is an element of $\mathcal{C}_z$.

For this code, the dispersive interaction lasts for
\begin{equation}
t_{\pi}=f(\chi,k)=\frac{\pi}{2\chi}
\end{equation}
 
\subsection{Dual-rail Qubit}
A single logical qubit in the Dual-rail qubit representation is illustrated by a pair of modes,  one excitation in each mode \cite{Chuang1995}
\begin{equation}
|0\rangle_L=|01\rangle,\quad |1\rangle_L=|10\rangle
\end{equation}
This qubit has been explored in superconducting microwave cavities, where its $Z(\theta)$ rotation is implemented by a beam-splitter interaction between cavities\cite{Teoh2023}.

without loss of generality, we apply the protocol to the first mode and obtain the greatest common divisor
\begin{equation}
k=\gcd\{0,1\}=1
\end{equation}
This leads to opposite parity between the logical codewords; This implies that the dual-rail qubit is a member of $\mathcal{C}_z$.

The dispersive interaction time becomes
\begin{equation}
t_{\pi}=f(\chi,k)=\frac{\pi}{\chi}
\end{equation}

\subsection{Wasilewski-Banaszek Qubit}
Logical basis states of this qubit are represented by \cite{Wasilewski2007}
\begin{equation}
|0\rangle_L=\frac{1}{\sqrt{3}}(|003\rangle+|030\rangle+|300\rangle),\quad |1\rangle_L=|111\rangle
\end{equation}
This qubit is still unexplored in circuit QED. So, we present this example to show the applicability of our protocol for future works.

By calculating the greatest common divisor
\begin{equation}
k=\gcd\{0,1,3\}=1
\end{equation}
and applying the protocol only to a single mode results in nonuniform parity within the logical $|0\rangle_L$. Furthermore, this outcome implies that we can apply the single-mode class condition however simultaneously the protocol produces a relative phase among states in the logical $|0\rangle_L$.
\begin{equation}
|\psi\rangle\propto\frac{1}{\sqrt{3}}(|003\rangle+|030\rangle+e^{i\theta}|300\rangle)+e^{i\theta}|111\rangle
\end{equation}
this issue can be overcome, provided that the protocol is applied consistently across the modes. In other words, each mode must be coupled to its own ancilla. The only difference is that we need to apply $U_4$ with $\theta=\frac{\theta}{4}$ to reach a consistent result with the protocol
\begin{multline}
|\psi\rangle=\frac{1}{\sqrt{3}}(e^{-i\theta/4}|003\rangle+e^{-i\theta/4}|030\rangle+e^{-i\theta/4}|300\rangle)\\+e^{i3\theta/4}|111\rangle\\
\end{multline}
which is equivalent to
\begin{equation}
|\psi\rangle\propto\frac{1}{\sqrt{3}}(|003\rangle+|030\rangle+|300\rangle)+e^{i\theta}|111\rangle
\end{equation}
therefore, the dispersive interaction time will be
\begin{equation}
t_{\pi}=f(\chi,k)=\frac{\pi}{\chi}
\end{equation}

\subsection{Two-Mode Fock-State Qubit}
For this encoding, the logical basis states are defined by \cite{Aydin2026}
\begin{multline}
\quad |0\rangle_L= \sqrt{\frac{3}{10}}|0,7\rangle+\sqrt{\frac{7}{10}}|5,2\rangle,\\ |1\rangle_L= \sqrt{\frac{7}{10}}|2,5\rangle-\sqrt{\frac{3}{10}}|7,0\rangle\\
\end{multline}
This is another example of unexplored qubits in cQED. Applying the protocol only to one mode, we get the greatest common divisor
\begin{equation}
k=\gcd(0,2,5,7)=1
\end{equation}
 however, this leads to opposite parity among components of each logical codeword. In contrast to the Wasilewski-Banaszek qubit, applying the protocol to both modes will not tackle the problem because both logical codewords obtain the same phase. So, for this case the protocol fails.


\section{CONCLUSIONS}

We have shown that a unified protocol can be used to implement the $Z(\theta)$ operation in a class of bosonic cQED qubits. The central observation of this work is that physical implementations of the same logical operation can appear substantially different for different bosonic Fock-based cQED qubits, while nevertheless sharing a common underlying structure. By identifying the conditions that captures this structure, we have classified bosonic cQED qubits that can benefit from a common protocol for the implementation of arbitrary $Z$ rotations.

We have illustrated the generality of the protocol through representative bosonic Fock-based cQED qubits. Although the $Z(\theta)$ operation in successful examples can be realized using distinct physical strategies, they can be described within the same protocol. This demonstrates that the protocol is not tied to a particular encoding or physical implementation, but instead captures a common feature shared by a class of bosonic cQED qubits.

Beyond providing a unified description of existing implementations, the conditions underlying the protocol offers a systematic way - a predictive power - to determine whether a new given bosonic Fock-based cQED qubit can be treated within the same framework. Thus, rather than requiring an encoding-specific construction of the $Z(\theta)$ operation, one can first test the relevant conditions and, when it is satisfied, apply the protocol.

\section{Acknowledgments}
We acknowledge helpful discussions with Varinder Singh.



\addcontentsline{toc}{chapter}{Bibliography}
\begin{thebibliography}{99}


   
    \bibitem{Deutsch1992}
    Rapid solutions of problems by quantum computations,
    David Deutsch and Richard Jozsa,
    Proceeding of the Royal Society of London A \textbf{439}, p. 553-558 (1992).
   
    \bibitem{Grover1996}
    A fast quantum mechanical algorithm for database search,
    Lov K. Grover,
    Proceedings, 28th Annual ACM Symposium on the Theory of computing, p. 212-219 (1996).
   
    \bibitem{Bennett1997}
    Strengths and weaknesses of quantum computing,
    Charles H. Bennett, Ethan Bernstein, Gilles Brassard, and Umesh Vazirani,
    SIAM Journal on Computing \textbf{26}, p. 1510-1523 (1997).
   
   
    \bibitem{Shor1999}
    Polynomial-Time Algorithms for Prime Factorization and Discrete Logarithms on a Quantum Computer,
    Peter W. Shor,
    SIAM Review, \textbf{41}, 303-332 (1999).
   
   
    \bibitem{Cirac1995}
    Quantum computations with cold trapped ions,
    J. I. Cirac and P. Zoller,
    Phys. Rev. Lett. \textbf{74}, 4091 (1995).
   
    \bibitem{Leibfried2003}
    Quantum dynamics of single trapped ions,
    D. Leibfried, R. Blatt, C. Monroe, and D. Wineland,
    Rev. Mod. Phys. \textbf{75}, 281 (2003).
   
    \bibitem{Haffner2008}
    Quantum computing with trapped ions,
    H. Haffner, C. F. Roos, and R. Blatt
    Physics Reports \textbf{469}, 155-203 (2008).
   
    \bibitem{Blatt2012}
    Quantum simulations with trapped ions,
    R. Blatt and C. F. Roos ,
    Nature Physics 8, 277-284 (2012).
   

   \bibitem{Bloch2008}
    Many-body physics with ultracold gases,
    Immanuel Bloch, Jean Dalibard, and Wilhelm Zwerger,
    Rev. Mod. Phys. \textbf{80}, 885 (2008).
   
   
    \bibitem{Saffman2010}
    Quantum information with rydberg atoms,
    M. Saffman, T. G. Walker, and K. Molmer,
    Rev. Mod. Phys. \textbf{82}, 2313 (2010).
   
    \bibitem{Henriet2020}
    Quantum computing with neutral atoms,
    Loic Henriet, Lucas Beguin, Adrien Signoles, Thierry Lahaye, Antonie Browaeys, Georges-Olivier Reymond, Christophe Jurczak,
    Quantum \textbf{4}, 327 (2020).
   
    \bibitem{Kaufman2021}
    Quantum science with optical tweezer arrays of ultracold atoms and molecules,
    A. M. Kaufman, K. K. Ni,
    Nature Physics \textbf{17}, 1324-1333 (2021).
   
   
   
   
    \bibitem{Blais2021}
    Circuit quantum electrodynamics,
    Alexandre Blais, Arne L. Grimsmo, S. M. Girvin, and Andreas Wallraff,
    Rev. Mod. Phys. \textbf{93}, 025005 (2021).
   
    \bibitem{Blais2004}
    Cavity quantum electrodynamics for superconducting electrical circuits: An architecture for quantum computation,
    Alexandre Blais, Ren-Shou Huang, Andreas Wallraff, S. M. Girvin, and R. J. Schoelkopf,
    Phys. Rev. A \textbf{69}, 062320 (2004).
   
    \bibitem{Vion2002}
    Manipulating the quantum state of an electrical circuit,
    D. Vion, A. Aassime, A. Cottet, P. Joyez, H. Pothier, C. Urbina, D. Esteve, M. H. Devoret,  
    Science \textbf{296}, p. 886-889, (2002)
   
    \bibitem{Koch2007}
    Charge-insensitive qubit design derived from the Cooper pair box,
    Jens Koch, Terri M. Yu, Jay Gambetta, A. A. Houck, D. I. Schuster, J. Majer, Alexander Blais, M. H. Devoret, S. M. Girvin, and R. J. Schoelkopf,
    Phys. Rev. A\textbf{76}, 042319 (2007).
   
    \bibitem{Schreier2008}
    Suppressing charge noise decoherence in superconducting charge qubits,
    J. A. Schreier, A. A. Houck, Jens Koch, D. I. Schuster, B. R. Johnson, J. M. Chow, J. M. Gambetta, J. Majer, L. Frunzio, M. H. Devoret, S. M. Girvin, and R. J. Schoelkopf,
    Phys. Rev. B \textbf{77}, 180502(R) (2008).
   
    \bibitem{Leghtas2013}
    Hardware-Efficient Autonomous Quantum Memory Protection,
    Z. Leghtas, G. Kirchmair, B. Vlastakis, R. J. Schoelkopf, M. H. Devoret and M. Mirrahimi,
    Phys. Rev. Lett. \textbf{111}, 120501 (2013).
   
    \bibitem{Terhal2016}
    Encoding a qubit into a cavity mode in circuit QED using phase estimation,
    B. M. Terhal and D. Weigand,
    Phys. Rev. A \textbf{93}, 012315 (2016).
   
    \bibitem{Ofek2016}
    Extending the lifetime of a quantum bit with error correction in superconducting circuits,
    N. Ofek, A. Petrenko, R. Heeres, P. Reinhold, Z. Leghtas, B. Vlastakis, Y. Liu, L. Frunzio, S. M. Girvin, L. Jiang, M. Mirrahimi, M. H. Devoret and R. J. Schoelkopf, , Nature \textbf{536}, 441-445 (2016).

      \bibitem{Hu2019}
    Quantum error correction and universal gate set operation on a binomial bosonic logical qubit,
    L. Hu, Y. Ma, W. Cai, X. Mu, Y. Xu, W. Wang, Y. Wu, H. Wang, Y. P. Song, C. L.
    Zou, S. M. Girvin, L. M. Duan and L. Sun,
    Nature Physics \textbf{15}, p. 503–508 (2019).
   
   
    \bibitem{Ying2015}
    Resource Costs for Fault-Tolerant Linear Optical Quantum Computing,
    Ying Li, Peter C. Humphreys, Gabriel J. Mendoza, and Simon C. Benjamin,
    Phys. Rev. X \textbf{5}, 041007 (2015).
   
   
    \bibitem{Reagor2016}
    Quantum memory with millisecond coherence in circuit QED,
    Matthew Reagor, Wolfgang Pfaff, Christopher Axline, Reinier W. Heeres, Nissim Ofek, Katrina Sliwa, Eric Holland, Chen Wang, Jacob Blumoff, Kevin Chou, Michael J. Hatridge, Luigi Frunzio, Michel H. Devoret, Liang Jiang, and Robert J. Schoelkopf,
    Phys. Rev. B \textbf{94}, 014506 (2016).
   
    \bibitem{Rosenblum2018}
    Fault-tolerant detection of a quantum error,
    S. Rosenblum, P. Reinhold, M. Mirrahimi, Liang Jiang, L. Frunzio, R.J. Schoelkopf,
    Science \textbf{361}, p.266 (2018).
   
    \bibitem{Sivak2023}
    Real-time quantum error correction beyond break-even,
    V. V. Sivak, A. Eickbusch, B. Royer, S. Singh, I. Tsioutsios, S. Ganjam, A. Miano, B. L. Brock, A. Z. Ding, L. Frunzio, S. M. Girvin, R. J. Schoelkopf, M. H. Devoret,
    Nature \textbf{616}, p. 50-55 (2023).
   
   
     \bibitem{Mirrahimi2014}
    Dynamically protected cat-qubits: a new paradigm for universal quantum computation,
    Mazyar Mirrahimi, Zaki Leghtas, Victor V Albert, Steven Touzard, Robert J Schoelkopf, Liang Jiang and Michel H Devoret
    New J. Phys. 16 045014 (2014).
   
    \bibitem{Heeres2017}
    Implementing a universal gate set on a logical qubit encoded in an oscillator,
    Reinier W. Heeres, Philip Reinhold, Nissim Ofek, Luigi Frunzio, Liang Jiang, Michel H. Devoret, Robert J. Schoelkopf
    Nature Communications\textbf{ 8}, 94 (2017).
   
     \bibitem{Hu2019}
    Quantum error correction and universal gate set operation on a binomial bosonic logical qubit,
    L. Hu, Y. Ma, W. Cai, X. Mu, Y. Xu, W. Wang, Y. Wu, H. Wang, Y. P. Song, C.-L. Zou, S. M. Girvin, L-M. Duan , L. Sun
    Nature Physics \textbf{15} 503-508 (2019)
   
   
    \bibitem{Ma2020}
   Error-transparent operations on a logical qubit protected by quantum error correction,
   Y. Ma, Y. Xu, X. Mu, W. Cai, L. Hu, W. Wang, X. Pan, H. Wang, Y. P. Song, C.-L. Zou, L. Sun
   Nature Physics \textbf{16}, 827–831 (2020).
   
   \bibitem{Xu2022}
   Engineering fast bias-preserving gates on stabilized cat qubits,
  Qian Xu, Joseph K Iverson, Fernando G.S.L. Brandao, Liang Jiang
   Phys. Rev. Research \textbf{4}, 013082 (2022).
   
   
   
        \bibitem{Teoh2023}
    Dual-rail encoding with superconducting cavities,
    James D. Teoh , Patrick Winkel , Harshvardhan K. Babla, Benjamin J. Chapman , Jahan Claes , Stijn J. de Graaf , John W. O. Garmon, William D. Kalfus, Yao Lu, Aniket Maiti, Kaavya Sahay , Neel Thakur , Takahiro Tsunoda, Sophia H. Xue , Luigi Frunzio , Steven M. Girvin, Shruti Puri, and Robert J. Schoelkopf,
    Proceedings of the National Academy of Sciences \textbf{120}, e2221736120 (2023).
   
     
    \bibitem{Marquet2024}
    Autoparametric Resonance Extending the Bit-Flip Time of a Cat Qubit up to 0.3 s,
   Antoine Marquet, Antoine Essig, Joachim Cohen, Nathanaël Cottet, Anil Murani, Emanuele Albertinale, Simon Dupouy, Audrey Bienfait, Théau Peronnin, Sébastien Jezouin, Raphaël Lescanne, Benjamin Huard
    PHYSICAL REVIEW X \textbf{14}, 021019 (2024).
   
     \bibitem{Shaghaghi}
    Error-correctable CLY bosonic qubit in the four-photon subspace of two microwave cavities,
    Vahid Shaghaghi \textit{et al}, in preparation
   
     \bibitem{Gautier2023}
    Designing High-Fidelity Zeno Gates for Dissipative Cat Qubits,
    Ronan Gautier, Mazyar Mirrahimi, and Alain Sarlette,
    PRX Quantum \textbf{4}, 040316 (2023).
   
    \bibitem{Xu2023}
    Autonomous quantum error correction and fault-tolerant quantum computation with squeezed cat qubits,
    Qian Xu, Guo Zheng, Yu-Xin Wang, Peter Zoller, Aashish A. Clerk, Liang Jiang
    npj Quantum Information \textbf{9}, 78 (2023).
   
       \bibitem{Liu2026}
    Hybrid Oscillator-Qubit Quantum Processors: Instruction Set Architectures, Abstract Machine Models, and Applications,
    Yuan Liu, Shraddha Singh, Kevin C Smith, Eleanor Crane, John M Martyn, Alec Eickbusch, Alexander Schuckert, Richard D Li, Jasmine Sinanan-Singh, Micheline B Soley, Takahiro Tsunoda, Isaac L Chuang, Nathan Wiebe, Steven M Girvin
    PRX Quantum\textbf{7}, 010201 (2026).
   
   
    \bibitem{Cochrane1999}
    Macroscopically distinct quantum superposition states as a bosonic code for amplitude damping,
    P. T. Cochrane, G. J. Milburn, W. J. Munro,
    Physical review A \textbf{59}, 2631 (1999).
   
    \bibitem{Iyama2024}
     Observation and manipulation of quantum interference in a superconducting Kerr parametric oscillator,
    Daisuke Iyama, Takahiko Kamiya, Shiori Fujii, Hiroto Mukai, Yu Zhou, Toshiaki Nagase, Akiyoshi Tomonaga, Rui Wang, Jiao-Jiao Xue, Shohei Watabe, Sangil Kwon, Jaw-Shen Tsai
    Nature Communications \textbf{15}, 86 (2024).
   
    \bibitem{Michael2016}
    New Class of Quantum Error-Correcting Codes for a Bosonic Mode,
    M. H. Michael, M. Silveri, R. T. Brierley, V. V. Albert, J. Salmilehto, L. Jiang, and S. M. Girvin,  
    Physical review X \textbf{6}, 031006 (2016).
   
    \bibitem{Chuang1995}
    Simple quantum computer,
    I. L. Chuang, Y. Yamamoto,
    Physical review A \textbf{52}, 3489 (1995).
   
   
    \bibitem{Wasilewski2007}
    Protecting an optical qubit against photon loss,
    W. Wasilewski and K. Banaszek,
    Physical review A \textbf{75}, 3489 (2007).
   
    \bibitem{Aydin2026}
    Quantum error correction beyond SU(2): spin, bosonic, and permutation-invariant codes from convex geometry,
    Arda Aydin, Victor V. Albert, Alexander Barg,
    PRX QUANTUM \textbf{7}, 010341 (2026)





\end{thebibliography}
\end{document}